\documentclass[preprint,twocolumn]{ceurart}

\usepackage[english]{babel}
\usepackage{microtype}
\usepackage{hyperref}

\conference{Applications of Generative AI to support teaching and learning in Higher Education, co-located with AIED 2025. Palermo Italy}

\title{Do AI tutors empower or enslave learners? Toward a critical use of AI in education}

\author[1]{Lucile Favero}[%
orcid=0009-0005-2981-0124,
email=lucile@ellisalicante.org
]
\cormark[1]
\fnmark[1]
\cortext[1]{Corresponding author: Lucile Favero lucile@ellisalicante.org}

\author[2]{Juan Antonio Pérez-Ortiz}[%
orcid=0000-0001-7659-8908,
]
\cormark[2]
\fnmark[2]

\author[3]{Tanja Käser}[%
orcid=0000-0003-0672-0415,
]
\cormark[3]
\fnmark[3]

\author[1]{Nuria Oliver}[%
orcid=0000-0001-5985-691X,
]
\cormark[1]
\fnmark[1]
\address[1]{ELLIS Alicante, Spain}
\address[2]{Universitat d'Alacant, Spain}
\address[3]{École Polytechnique Fédérale de Lausanne, EPFL, Switzerland}

\begin{document}
\maketitle

\begin{abstract}
The increasing integration of AI tools in education presents both opportunities and challenges, particularly regarding the development of the students' critical thinking skills. This position paper argues that while AI can support learning, its unchecked use may lead to cognitive atrophy, loss of agency,  emotional risks, and ethical concerns,  ultimately undermining the core goals of education. Drawing on cognitive science and pedagogy, the paper explores how over-reliance on AI can disrupt meaningful learning, foster dependency and conformity, undermine the students’ self-efficacy, academic integrity, and well-being, and raise concerns about questionable privacy practices. It also highlights the importance of considering the students' perspectives and proposes actionable strategies to ensure that AI serves as a meaningful support rather than a cognitive shortcut. The paper advocates for an intentional, transparent, and critically informed use of AI that empowers rather than diminishes the learner.
\end{abstract}

\section{Introduction}

\begin{quote}
\emph{“\textbf{Sapere aude}! Have courage to use your own reason!”} \\
\hfill --- Immanuel Kant, \textit{An Answer to the Question: What is Enlightenment?} (1784)
\end{quote}

Artificial Intelligence (AI) is increasingly recognized as a transformative tool for education. According to UNESCO, AI has the potential to tackle some of the most pressing educational challenges, enhance teaching and learning practices, and accelerate progress toward the Sustainable Development Goal 4 (SDG 4) \cite{unescoAIeducation}. Yet, the rapid development and unprecedented adoption rates --particularly of chatbots-- present significant risks that often outpace existing policy and regulatory frameworks.

Empirical evidence supports the educational benefits of AI-driven systems. Intelligent Tutoring Systems, for instance, have shown \textbf{generally positive effects }on student learning and performance \cite{letourneauSystematicReviewAIdriven2025}. Generative AI could also serve as a personal tutor—particularly valuable in contexts of\textbf{ teacher shortages}—and produce adaptive, personalized, diverse, and \textbf{up-to-date educational materials} tailored to individual learners’ needs \cite{tafazoliExploringPotentialGenerative2024, al-zahraniExploringImpactArtificial2024}. These systems aim to promote\textbf{ self-paced, autonomous learning} by supporting the key phases of \textbf{self-regulated learning}: goal setting, performance monitoring, and reflection \cite{lanQualitativeSystematicReview2025, vorobyevaPersonalizedLearningAI2025,al-zahraniExploringImpactArtificial2024}. \textbf{Instant feedback and 24/7 availability } should support the learner’s ability to progress independently and at their own pace \cite{vorobyevaPersonalizedLearningAI2025}. Consequently, AI has the potential to reduce teacher workload, increase cost-effectiveness, and personalize and scale educational access, contributing to a more \textbf{equitable and democratized} learning environment \cite{worldbank2024ai}. 

\begin{figure}[ht]
\centering
\includegraphics[width=0.47\textwidth]{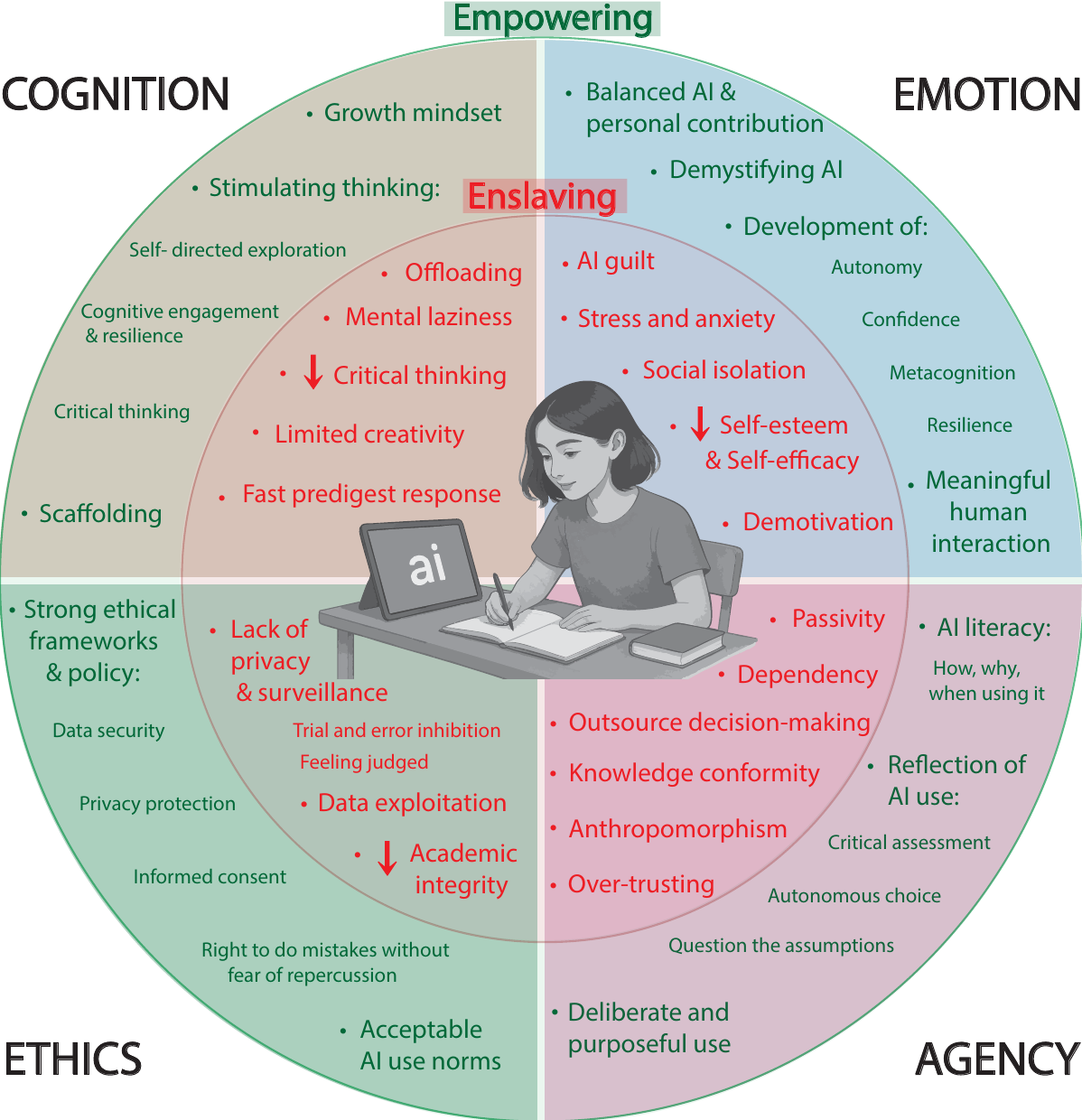}
\caption{Ensuring a Human-Centered, responsible, and critical use of AI in education: challenges and interventions}
\label{fig:visual}
\end{figure}

Despite its many advantages, the integration of AI in education raises significant concerns. 
\textbf{Over-reliance on AI} may lead to \textbf{reduced critical thinking}, as students increasingly offload cognitive effort \cite{ravseljHigherEducationStudents2025,gerlichAIToolsSociety2025a}, and a \textbf{loss of agency}, where learners feel less in control of their own learning processes \cite{darvishi2024impact,roe2024generative}. 
AI use may have a \textbf{significant negative impact on the students’ emotional well-being} by leading to dependency, diminishing self-efficacy and self-esteem, and triggering feelings such as anxiety, AI guilt,  impostor syndrome, cognitive dissonance, or digital isolation \cite{rodriguez-ruizArtificialIntelligenceUse2025}. \textbf{Data privacy} is another concern as student data could potentially be misused or inadequately protected make \cite{kwapiszPrivacyConcernsStudent2024}.

In this paper, we reflect on and discuss these ambivalent implications of AI integration in education, with a focus on the impact of chatbots on critical thinking and learner autonomy by analyzing: cognitive atrophy in Section \ref{sec:co}, loss of agency in Section \ref{sec:agency}, undermined emotional well-being in Section \ref{sec:emo}, and ethical and academic concerns in Section \ref{sec:ethic}. The necessary student perspectives are presented in Section \ref{sec:stu}. Finally, we propose in Section \ref{sec:action} three additional principles for a responsible, constructive, and empowering use of AI in learning environments. Figure~\ref{fig:visual} provides an overview of the challenges and proposed interventions. 

In sum, we argue for a human-centic approach, such that the \textbf{integration of AI in education prioritizes the learners' growth, autonomy, and well-being over automation}. 

\section{Cognitive atrophy of the learning} \label{sec:co}
The widespread adoption of AI tools, especially chatbots, in education raises concerns about how students engage with information. Rather than critically processing content, many users tend to accept AI-generated responses quickly and uncritically, preferring fast, simplified answers over independent reasoning \cite{zhaiEffectsOverrelianceAI2024, ravseljHigherEducationStudents2025,williams2024ethical,letourneauSystematicReviewAIdriven2025, zhangYouHaveAI2024}.

\subsection{Cognitive offloading and critical thinking}
\textbf{Cognitive offloading}, \emph{i.e.}, the use of external tools to reduce the mental load on working memory, is a well-documented phenomenon in education \cite{riskoCognitiveOffloading2016,gerlichAIToolsSociety2025a}. Tools like calculators, gestures, or diagrams can aid learning when they offload non-essential demands and free cognitive resources for deeper thinking. However, not all offloading is beneficial. If the mental effort that is freed is not redirected towards other meaningful cognitive tasks, it can undermine learning outcomes \cite{riskoCognitiveOffloading2016}.

Certain use of AI has been found to contribute to the offloading of core thinking tasks, including analytical, reasoning, or synthesis tasks, weakening the students' critical engagement with the content \cite{zhaiEffectsOverrelianceAI2024, gerlichAIToolsSociety2025a,ravseljHigherEducationStudents2025,williams2024ethical,letourneauSystematicReviewAIdriven2025, zhangYouHaveAI2024}. Over-reliance on AI may lead to reduced independent and critical thinking, memory, creativity, deep reasoning, and the motivation to exert cognitive effort. A user study by \citet{gerlichAIToolsSociety2025a} across age and education levels found that a higher reliance on AI was correlated with lower performance on the HCTA critical thinking assessment by Halpern \cite{halpern2010halpern} mediated by cognitive offloading. The study emphasizes that automation of analytical tasks can erode the learners' ability for independent reasoning, complex problem-solving, and active learning.

These effects are especially concerning in educational settings, where cultivating intellectual resilience and independent thought is essential \cite{mezirowTransformativeLearningTheory1997, letourneauSystematicReviewAIdriven2025}.
Integrating AI in education must be balanced with deliberate strategies to preserve and promote critical thinking. \textbf{Fostering cognitive resilience instead of automation} alone should be a key goal of AI-enhanced education.

\subsection{Pedagogical and Cog-Sci Principles}
Effective learning is grounded in well-established principles from pedagogy and cognitive science.
\textbf{Active learning} seeks to engage learners in activities that require analysis, synthesis, and evaluation of content, rather than passive reception. Empirical evidence demonstrates that this approach leads to significantly better outcomes than traditional, more passive, instruction methods \cite{freeman2014active}.
\textbf{Constructivism} encourages learners to actively construct knowledge by integrating new information with their prior understanding, through personal experience and social interaction. This dynamic process not only fosters deeper understanding, but also cultivates transferable skills \cite{fosnot2013constructivism}.
\textbf{Scaffolding} refers to the gradual reduction of instructional support as learners gain competence. This method promotes the autonomy of the student: by the end of the process, the students are able to solve complex problems independently \cite{van2002scaffolding}.

Unfortunately, most existing chatbots that are heavily used by students do not embody these principles \cite{letourneauSystematicReviewAIdriven2025,openai2025aiready,ravseljHigherEducationStudents2025}. They typically deliver pre-digested information in a static, directive manner, limiting the learner’s opportunity to engage in critical thinking or self-directed exploration. This passive interaction can hinder learning outcomes \cite{williams2024ethical,zhaiEffectsOverrelianceAI2024, gerlichAIToolsSociety2025a}. In contrast, AI-powered tutoring systems that are explicitly designed according to educational theory show considerable promise. Well-designed AI tutors can outperform even active learning techniques in rigorous academic settings \cite{kestin*AITutoringOutperforms2024}. Students using these AI tutors learned more than twice and in less time than those engaged in traditional active learning, highlighting both the efficiency and effectiveness of pedagogically-informed AI systems.

These results emphasize a critical distinction: AI becomes truly transformative only when aligned with cognitive and educational science. The most effective systems integrate scaffolding, manage cognitive load, and foster a growth mindset, three foundational principles in educational psychology. By contrast, generic generative AI tools often act as shortcuts to answers, outsourcing rather than stimulating the thinking process \cite{gerlichAIToolsSociety2025a,al-zahraniUnveilingShadowsHype2024}. This distinction matters. \textbf{When AI supports, rather than replaces} the cognitive engagement of the learner, it can meaningfully foster critical thinking, autonomy, and deep learning.

\subsection{Superficial learning vs. desirable difficulties}
Effective learning entails mental friction, \emph{i.e.}, it requires an effort but also leads to deeper understanding and long-term retention \cite{bjork2011making}. Research on desirable difficulties highlights strategies like spacing, interleaving, retrieval practice, and delayed feedback as essential for robust learning, even if they momentarily slow performance. These methods promote knowledge transfer and critical thinking precisely because they are mentally taxing \cite{de2023worth,meraUnravelingBenefitsExperiencing2022}.

However, learners often try to avoid such an effort. Studies show that high perceived difficulty and low short-term performance can discourage engagement, despite clear long-term benefits. Interestingly, \citet{deslauriersMeasuringActualLearning2019} found that students in active learning environments learned more but felt they learned less. Mental effort was misinterpreted as failure, while smooth lectures mistakenly felt more effective, though they were not. This cognitive bias leads students to favor fluent, low-effort activities that give the illusion of learning, such as re-reading polished explanations, without fostering deep processing \cite{de2023worth,biwer2020future,  zhangYouHaveAI2024}. 

AI tools, and particularly chatbots, may exacerbate this issue. By offering quick, fluent, and simplified answers, they reduce the cognitive struggle which is essential to learning \cite{zhaiEffectsOverrelianceAI2024}. Their convenience may lead to passive consumption, decreased research and reasoning skills, and a growing dependence on pre-digested knowledge \cite{howladerFactorsInfluencingAcceptance2025}. Ease of use is appealing, but true learning comes from effort, complexity, and time.

To leverage AI without undermining learning, its use must be aligned with the principles of \textbf{desirable difficulty}. Tools should preserve cognitive effort, prompt active engagement, and help learners understand that struggle is not a sign of failure and it is necessary for the learning itself \cite{bjork2011making,biwer2020future,meraUnravelingBenefitsExperiencing2022}.

\section{Loss of agency}\label{sec:agency}
While AI tools offer substantial support, their increasing presence in educational contexts may foster over-reliance, undermining students' ability to engage deeply and reflect critically \cite{al-zahraniUnveilingShadowsHype2024}. This dependence becomes particularly dangerous when AI-generated content is inconsistent, biased, false or misleading, potentially reinforcing inequalities and spreading false or fabricated information.

\subsection{Dependency}
As students increasingly turn to AI for academic support, particularly for feedback, they may do so at the expense of developing essential cognitive and self-regulatory skills. When AI assistance is withdrawn, performance tends to decline, revealing a lack of internalized learning strategies and self-monitoring habits \cite{darvishi2024impact}. Even when AI is used in conjunction with metacognitive scaffolds, students often fail to regain full autonomy.

This growing reliance on AI raises concerns about its deeper cognitive and educational implications. Routine use of AI tools can hinder the development of metacognitive skills, independent thinking, and intellectual agency \cite{lanQualitativeSystematicReview2025, rodriguez-ruizArtificialIntelligenceUse2025}. As students begin to outsource decision-making to the machine, they risk becoming passive recipients of information rather than critical participants in the learning process \cite{al-zahraniExploringImpactArtificial2024}. This passivity is linked to broader deficits, including reduced creativity, increased mental laziness, and diminished capacity for critical thought \cite{zhangYouHaveAI2024}. Moreover, dependency on AI tools can lead to the uncritical acceptance of their outputs. When students perceive these systems as convenient, accurate, and reliable, they may stop questioning the information provided, which fosters cognitive dependency, \emph{i.e.}, the erosion of the ability to assess, verify, and challenge content independently \cite{gerlichAIToolsSociety2025a}. Finally, excessive use of AI can deteriorate emotional intelligence and human connection, contributing to isolation and anxiety \cite{klimova2025exploring}.

These risks are especially pronounced among students with lower digital literacy. These learners are more susceptible to misinformation and over-reliance on AI systems, further exacerbating educational inequalities \cite{becirovic2025exploring,zhangYouHaveAI2024}. As AI-generated prompts increasingly shape learning trajectories, the roles of the students are shifting from active knowledge constructors to passive respondents, which threatens both individual learner autonomy and the foundational principles of educational equity \cite{roe2024generative}.

To mitigate these AI-dependent behaviors, it is essential to develop the skill of a \textbf{deliberate and purposeful use of AI} \cite{zhangYouHaveAI2024}.

\subsection{Over-trusting}
This AI dependence is exacerbated by an apparent reliability of AI tools. As a result, the learners' capacity for independent judgment is eroded, and their critical engagement with learning content is weakened. \cite{gerlichPowerVirtualInfluencers2023}.
The human-like nature of many AI systems compounds the issue. Students often accept AI outputs as accurate simply because they are presented in a conversational, articulate and somewhat authoritative tone \cite{schaaff2024impacts}. This perceived credibility fosters uncritical trust, weakening students’ judgment and reinforcing the notion that AI “knows better.” Over time, this trust can lead to the outsourcing of cognitive tasks: students stop thinking for themselves and defer to the machine. Such trust reduces the impulse to question information, directly undermining critical thinking \cite{gerlichPowerVirtualInfluencers2023}. The more students rely on AI, the less they engage in evaluative thought, leading to cognitive offloading and a decline in deep learning. Chatbots, in particular, have been shown to erode student confidence and reduce autonomous decision-making \cite{williams2024ethical}. 

To reduce overtrust in AI, \textbf{students should be taught how AI systems actually work}, highlighting that they are algorithms trained on data, not conscious beings. This understanding helps demystify AI, fosters critical thinking, and restores student agency by encouraging them to question rather than blindly trust AI outputs \cite{bastaniGenerativeAICan2024}.

\subsection{Conformity}
 By providing ready-made answers or persuasive feedback, chatbots can limit creative exploration and encourage intellectual conformity \cite{tafazoliExploringPotentialGenerative2024}. Rather than promoting diverse perspectives, it risks promoting uniform responses, homogenizing the richness of student thought \cite{al-zahraniExploringImpactArtificial2024}. Indeed, AI systems may subtly shape the way students think.

 Meaningful learning is inseparable from agency. \textbf{Transformative learning requires the ability to think critically, question assumptions, and make autonomous choices} \cite{mezirowTransformativeLearningTheory1997}. Without these capacities, intellectual growth is stunted, and the very purpose of education is compromised.

\section{Undermined emotional well-being}\label{sec:emo}
Beyond cognitive and agency effects, the integration of AI tools into education raises critical concerns about the learners' emotional well-being and identity formation.

\subsection{Self-efficacy and self-esteem}
Students with low academic self-efficacy or self-esteem are more likely to rely on AI to compensate for what they see as their own shortcomings \cite{lanQualitativeSystematicReview2025, rodriguez-ruizArtificialIntelligenceUse2025, zhangYouHaveAI2024}. This reliance can create a harmful cycle: the more students use AI to avoid academic challenges, the less confident they become in their own abilities. This loss of confidence reduces their willingness to take initiative, which in turn increases dependence on AI and further weakens self-belief \cite{williams2024ethical, chanExploringFactorsAI2024}. Such students also tend to feel more stress and face unrealistic academic expectations, which pushes them even more toward the use of AI. As a result, their ability to think critically, be creative, and learn independently may decline over time \cite{zhangYouHaveAI2024}.

When students see AI as faster and more capable than themselves, they may begin to undervalue their own efforts and knowledge. One student said, \emph{“I will never be better than AI”} \cite{lanQualitativeSystematicReview2025}, illustrating how AI can unintentionally lower the students’ motivation and belief in their potential, leading to the \emph{Impostor Syndrome} \cite{chanExploringFactorsAI2024}. Students who have a better understanding of how AI systems work, including what they can and cannot do, show higher levels of academic self-efficacy \cite{becirovic2025exploring}. Thus, \textbf{demystifying AI} will contribute to support the students' trust in their own abilities.

\subsection{AI guilt and cognitive dissonance}

Relying on AI for academic work can lead to \emph{AI guilt}: feelings of shame, anxiety, and moral discomfort tied to the use of AI tools \cite{chanExploringFactorsAI2024}. Students express sentiments like: \emph{“I feel like I am not being truthful when I use it,”} and describe feeling “lazy” or afraid of being judged by peers and instructors \cite{chanExploringFactorsAI2024}. These emotions affect not only their well-being but also their sense of identity, self-worth and personal agency.

Such feelings often lead to \emph{cognitive dissonance}, \emph{i.e.}, the psychological discomfort that occurs when actions conflict with deeply held beliefs \cite{elliot1994motivational}. Cognitive dissonance helps explain the tension students feel when they value originality and personal effort, yet use AI tools that may undermine these ideals. For instance, a student may feel proud of an AI-assisted essay but also guilty that it does not reflect their own independent thinking \cite{chanExploringFactorsAI2024}. This internal conflict can be intense. When students believe that genuine academic work should come from human creativity and effort, using AI challenges their core values. The result is often stress, anxiety, and a weakened sense of authenticity in their learning journey \cite{lanQualitativeSystematicReview2025}.

To address feelings of guilt associated with using AI, it is important to \textbf{strike a balance between technological assistance and personal contribution}. Educators can promote this balance by encouraging students to use AI for guidance or support, while making sure that the final work genuinely represents the student’s own understanding and effort \cite{chanExploringFactorsAI2024}.

On the other hand, \emph{AI entitlement} reflects a growing belief that access to AI assistance is not only acceptable but a rightful part of modern education. The tension between these two concepts (guilt vs entitlement) reveals a broader cultural tension on how we define and value effort, ownership and fairness in learning and underscores the urgent need for clear norms about the use of AI in education. 

\subsection{Emotional risks}
Constant use of AI can contribute to technostress, digital fatigue, and social isolation, particularly when it displaces interpersonal interaction and collaborative learning \cite{klimova2025exploring}.
These emotional risks of AI use in education are not marginal and directly affect the motivation, identity, and well-being of the learner \cite{al-zahraniExploringImpactArtificial2024}. Designing pedagogical approaches that acknowledge and mitigate these risks is essential to ensure that \textbf{AI supports, rather than replaces,} the development of resilient, confident, and autonomous learners.

\section{Ethical and Academic concerns}\label{sec:ethic}
The integration of AI, especially chatbots, into educational environments raises profound ethical challenges, especially concerning student privacy, surveillance, academic integrity and the dynamics of power in digital learning spaces \cite{williams2024ethical, kwapiszPrivacyConcernsStudent2024}. 

\subsection{Privacy}
While AI educational technologies promise personalized support and enhanced learning outcomes, they also risk creating environments where students feel constantly observed, judged, and recorded \cite{al-zahraniUnveilingShadowsHype2024,shores2024surveillance}. Constant tracking has psychological impacts: when students know that everything they write is potentially stored and analyzed, the natural process of learning through trial and error may be inhibited \cite{shores2024surveillance}. Making mistakes is an essential part of developing critical thinking \cite{mason2016learning,meraUnravelingBenefitsExperiencing2022}, yet, fear of doing something wrong can lead to shame, loss of confidence, and reduced willingness to take intellectual risks \cite{kwapiszPrivacyConcernsStudent2024,shores2024surveillance}, which goes against the goals of learning which should foster curiosity, resilience, and reflective thinking \cite{mezirowTransformativeLearningTheory1997}.

\subsection{Data exploitation}
These harms are compounded by the broader issue of data privacy \cite{klimova2025exploring}. Generative AI systems often manage sensitive student information, and their deployment raises serious concerns about how this data is stored, used, and protected \cite{al-zahraniExploringImpactArtificial2024}. The improper use or insecure storage of student data, ranging from performance metrics to behavioral patterns, poses substantial risks \cite{al-zahraniUnveilingShadowsHype2024}.

Although compliance with data protection laws, such as the European General Data Protection Regulation (GDPR), is essential, it is frequently difficult to enforce in practice \cite{williams2024ethical, howladerFactorsInfluencingAcceptance2025}. This difficulty is exacerbated by the power imbalance between educational institutions and students, who often have little understanding or control over what data is collected and how it is used. As a result, what is framed as educational support can easily become a justification for constant surveillance and data extraction \cite{kwapiszPrivacyConcernsStudent2024}.

We emphasize the need for strong ethical frameworks that comply with informed consent and the protection of the students’ fundamental rights, including the right to make mistakes without fear of judgment or repercussion. \textbf{AI in education should empower learners, not reduce them to data points.} Ensuring privacy and data security is not only a legal obligation but a moral imperative if AI is to support, rather than undermine, the development of critical and autonomous thinkers.
\vspace*{-0.15cm}
\subsection{Academic integrity}

When classroom environments lack pedagogical support or clear guidance, students are more likely to turn to generic AI chatbots and other AI tools, which serve as quick fixes rather than learning companions. This pattern of use fuels procrastination and ultimately undermines academic integrity \cite{niloyAIChatbotsDisguised2024}. Instead of fostering deep learning and critical engagement, poorly integrated AI tools in education encourage shortcut-seeking behaviors, ranging from superficial answers to plagiarism and cheating. The sophisticated language generation capabilities of today's chatbots challenge traditional plagiarism detection methods by producing text that is both original in structure and closely aligned with human writing patterns, making it difficult to distinguish between genuinely authored content and AI-assisted compositions using conventional tools. Without careful instructional design and clear boundaries for AI use, these technologies risk becoming enablers of academic misconduct rather than tools for intellectual growth \cite{williams2024ethical}.

In response to these challenges, institutions must approach AI integration with both ambition and caution. Effective use of AI in education requires a commitment to pedagogy, not just technology. At the same time, safeguards must be implemented to preserve academic integrity and support the development of honest, independent learners. This includes \textbf{designing assessments that reward critical engagement, embedding AI literacy in curricula, and establishing clear norms around acceptable AI use} \cite{al-zahraniUnveilingShadowsHype2024,kwapiszPrivacyConcernsStudent2024}.
\vspace*{-.15cm}
\section{Student perspectives}\label{sec:stu}
In support of a human-centered approach to the integration of AI in education, it is crucial to give voice to the students’ perspectives.
\subsection{AI adoption}
According to OpenAI, young adults aged 18–24 are the biggest adopters of ChatGPT in the U.S., with over one-third using the tool regularly \cite{openai2025aiready}. Among these users, more than a quarter of messages are related to education, including tutoring, writing support, and programming help. In a survey of 1,200 students in this age group, AI tools were most often used for initiating papers and projects (49\%), summarizing texts (48\%), brainstorming (45\%), topic exploration (44\%), and revising writing (44\%).

On a global scale, a 2024 international survey of 4,000 university students across 16 countries found that students value AI primarily for its timely support (63\%), help in understanding tool use (46\%), and access to training opportunities (45\%) \cite{digitaleducationcouncil2025}. 
\subsection{Concerns}
Students are aware of the potential negative consequences of their AI use and formulate their concerns: 61\% worry about data privacy, while many question the reliability of AI-generated content and the academic risks of over-reliance. They fear that excessive dependence on AI could undermine learning, critical thinking, and the instructional value of education \cite{digitaleducationcouncil2025}. Aligned with these findings, a qualitative study by \citet{gerlichAIToolsSociety2025a} found that students were increasingly aware of: (1) their dependency on AI for both routine and cognitive tasks; (2) the diminished opportunities for independent thinking; and (3) ethical issues such as bias, transparency, and decision-making influence. Table~\ref{tab:testi} provides selected participant exemplary comments regarding such concerns.

\begin{table}[ht]
\centering
\caption{Student concerns on their AI use \cite{gerlichAIToolsSociety2025a}. }
\begin{tabular}{|p{2.3cm}|p{4.6cm}|}
\hline
\textbf{Concerns} & \textbf{Testimonials} \\ \hline

High reliance on AI tools & 
\textit{“I use AI for everything, from scheduling to finding information. It's become a part of how I think.” “I can’t imagine functioning without my digital assistant.”} \\ \hline

Diminished cognitive engagement & 
\textit{"The more I use AI, the less I feel the need to problem-solve on my own. It's like I'm losing my ability to think critically.”} \\ \hline

Ethical implications (transparency, bias, influence) & 
\textit{“I sometimes wonder if AI is subtly nudging me toward decisions I wouldn't normally make.”,“I rarely reflect on the biases behind the AI recommendations; I tend to trust them outright.”} \\ \hline

\end{tabular}
\label{tab:testi}
\end{table}
\subsection{The educational level shapes AI critical use}
Attitudes toward AI use and the ability to critically evaluate its outputs vary significantly depending on one’s educational background. For instance, in \cite{gerlichAIToolsSociety2025a} a high schooler states: \emph{“I don't have the time or skills to verify what AI says; I just trust it.”} In contrast, a doctoral-level participant explained: \emph{“While I use AI tools regularly, I always make sure to critically evaluate the information I receive. My education has taught me the importance of not accepting things at face value, especially when it comes to AI, which can sometimes offer biased or incomplete information”}. These contrasting perspectives emphasize a pressing need for \textbf{inclusive AI literacy education} that equips all learners with the skills to engage critically with AI-generated content and avoid deepening cognitive and educational divides.

\section{Discussion}\label{sec:action}

It is evident that students should be taught to critically assess AI educational tools to recognize its limitations \cite{zhaiEffectsOverrelianceAI2024}. The risks of AI technology are not inherent to the tools themselves but lie in how they are utilized. Therefore, effective integration requires targeted educational, ethical, and infrastructural strategies. Beyond each end-of-section suggestion, we believe that the use of AI in education should:

\textbf{1. Empower reasoning.} Education should prioritize constructivist, active learning, and metacognitive strategies to reduce overreliance on AI and mitigate cognitive offloading  \cite{riskoCognitiveOffloading2016, niloyAIChatbotsDisguised2024}. For instance, rather than providing direct answers, AI tools should be used to support scaffolding, prompting deeper reflection, reasoning, and understanding \cite{letourneauSystematicReviewAIdriven2025, bastaniGenerativeAICan2024, favero2024enhancing}. Students must be encouraged to \textbf{engage their own reasoning} by working on tasks that require them to compare AI-generated outputs with human logic, question assumptions, and uncover blind spots \cite{zhaiEffectsOverrelianceAI2024, gerlichAIToolsSociety2025a}. 

\textbf{2. Foster emotional well-being.} To address feelings of guilt, confusion, or in-authenticity surrounding AI use, students need guidance not only on what AI can do, but also on \textbf{how and when to use it in ways that enhance learning}. Meta-cognitive training and critical thinking exercises are key to helping students reflect on their engagement with AI and build confidence in their intellectual abilities \cite{chanExploringFactorsAI2024}. We must emphasize that human value should not be measured against that of AI algorithms. 
AI systems lack consciousness and a lived experience in the physical world, rendering such comparisons fundamentally flawed and intellectually unproductive \cite{becirovic2025exploring}.
Furthermore, it is essential to preserve meaningful human interaction in education to maintain empathy, communication, and social learning \cite{al-zahraniUnveilingShadowsHype2024}.

\textbf{3. Institutional support} is necessary to achieve our vision of human-centric AI for education. First, \textbf{robust AI literacy programs} are needed to prepare students for responsible and informed use of AI tools \cite{becirovic2025exploring}, including data privacy, security, and algorithmic transparency, supported by audits and bias mitigation \cite{al-zahraniUnveilingShadowsHype2024}. As only 25\% of U.S. colleges currently offer formal AI training \cite{openai2025aiready}, there is a pressing need to develop AI-integrated curricula across all educational levels. Equally important is professional development for educators, enabling them to guide students in using AI ethically and effectively \cite{williams2024ethical}. Institutions must \textbf{define clear  policies} regarding (1) transparency and mitigation power dynamic \cite{kwapiszPrivacyConcernsStudent2024}, (2) data privacy that complies with informed consent and protection of the students' rights \cite{klimova2025exploring}, (3) acceptable AI use, plagiarism, and student accountability \cite{williams2024ethical}, and (4) inclusive and equitable access while preserving learner agency \cite{roe2024generative}. 
\vspace*{-.2cm}
\section{Conclusion}

AI offers unprecedented opportunities to support, democratize and personalize education. However, it also poses risks that need to be addressed before its potential is realized. In this paper, we have structured such risks in four key areas in the learning process: cognition, emotion, ethics and agency. We advocate for a reflective, interdisciplinary approach to AI in education that prioritizes pedagogical goals over mere technological capabilities.
Such an alignment needs to be continuously assessed, not assumed: if we aim to cultivate critical thinkers, we must critically evaluate the tools that we entrust to that help us with that task.

\section*{Acknowledgments}
This work has been funded by a nominal grant received at the ELLIS Unit Alicante Foundation from the Regional Government of Valencia in Spain (Convenio Singular signed with Generalitat Valenciana, Conselleria de Innovación, Industria, Comercio y Turismo, Dirección General de Innovación) and a grant by the Banc Sabadell Foundation.



\begin{thebibliography}{43}
\expandafter\ifx\csname natexlab\endcsname\relax\def\natexlab#1{#1}\fi
\providecommand{\url}[1]{\texttt{#1}}
\providecommand{\href}[2]{#2}
\providecommand{\path}[1]{#1}
\providecommand{\DOIprefix}{doi:}
\providecommand{\ArXivprefix}{arXiv:}
\providecommand{\URLprefix}{URL: }
\providecommand{\Pubmedprefix}{pmid:}
\providecommand{\doi}[1]{\href{http://dx.doi.org/#1}{\path{#1}}}
\providecommand{\Pubmed}[1]{\href{pmid:#1}{\path{#1}}}
\providecommand{\bibinfo}[2]{#2}
\ifx\xfnm\relax \def\xfnm[#1]{\unskip,\space#1}\fi
\bibitem[{{UNESCO}(2025)}]{unescoAIeducation}
\bibinfo{author}{{UNESCO}}, \bibinfo{title}{{Artificial intelligence in education}}, \bibinfo{howpublished}{\url{https://www.unesco.org/en/digital-education/artificial-intelligence}}, \bibinfo{year}{2025}. \bibinfo{note}{Accessed: 2025-05-25}.
\bibitem[{Létourneau et~al.(2025)Létourneau, Deslandes~Martineau, Charland, Karran, Boasen, and Léger}]{letourneauSystematicReviewAIdriven2025}
\bibinfo{author}{A.~Létourneau}, \bibinfo{author}{M.~Deslandes~Martineau}, \bibinfo{author}{P.~Charland}, \bibinfo{author}{J.~A. Karran}, \bibinfo{author}{J.~Boasen}, \bibinfo{author}{P.~M. Léger},
\newblock \bibinfo{title}{A systematic review of {{AI-driven}} intelligent tutoring systems ({{ITS}}) in {{K-12}} education},
\newblock \bibinfo{journal}{npj Science of Learning} \bibinfo{volume}{10} (\bibinfo{year}{2025}) \bibinfo{pages}{1--13}. \URLprefix \url{https://www.nature.com/articles/s41539-025-00320-7}. \DOIprefix\doi{10.1038/s41539-025-00320-7}.
\bibitem[{Tafazoli(2024)}]{tafazoliExploringPotentialGenerative2024}
\bibinfo{author}{D.~Tafazoli},
\newblock \bibinfo{title}{Exploring the potential of generative {{AI}} in democratizing {{English}} language education},
\newblock \bibinfo{journal}{Computers and Education: Artificial Intelligence} \bibinfo{volume}{7} (\bibinfo{year}{2024}) \bibinfo{pages}{100275}. \URLprefix \url{https://linkinghub.elsevier.com/retrieve/pii/S2666920X2400078X}. \DOIprefix\doi{10.1016/j.caeai.2024.100275}.
\bibitem[{Al-Zahrani and Alasmari(2024)}]{al-zahraniExploringImpactArtificial2024}
\bibinfo{author}{A.~M. Al-Zahrani}, \bibinfo{author}{T.~M. Alasmari},
\newblock \bibinfo{title}{Exploring the impact of artificial intelligence on higher education: {{The}} dynamics of ethical, social, and educational implications},
\newblock \bibinfo{journal}{Humanities and Social Sciences Communications} \bibinfo{volume}{11} (\bibinfo{year}{2024}) \bibinfo{pages}{1--12}. \URLprefix \url{https://www.nature.com/articles/s41599-024-03432-4}. \DOIprefix\doi{10.1057/s41599-024-03432-4}.
\bibitem[{Lan and Zhou(2025)}]{lanQualitativeSystematicReview2025}
\bibinfo{author}{M.~Lan}, \bibinfo{author}{X.~Zhou},
\newblock \bibinfo{title}{A qualitative systematic review on {{AI}} empowered self-regulated learning in higher education},
\newblock \bibinfo{journal}{npj Science of Learning} \bibinfo{volume}{10} (\bibinfo{year}{2025}) \bibinfo{pages}{1--16}. \URLprefix \url{https://www.nature.com/articles/s41539-025-00319-0}. \DOIprefix\doi{10.1038/s41539-025-00319-0}.
\bibitem[{Vorobyeva et~al.(2025)Vorobyeva, Belous, Savchenko, Smirnova, Nikitina, and Zhdanov}]{vorobyevaPersonalizedLearningAI2025}
\bibinfo{author}{K.~I. Vorobyeva}, \bibinfo{author}{S.~Belous}, \bibinfo{author}{N.~V. Savchenko}, \bibinfo{author}{L.~M. Smirnova}, \bibinfo{author}{S.~A. Nikitina}, \bibinfo{author}{S.~P. Zhdanov},
\newblock \bibinfo{title}{Personalized learning through {{AI}}: {{Pedagogical}} approaches and critical insights},
\newblock \bibinfo{journal}{Contemporary Educational Technology} \bibinfo{volume}{17} (\bibinfo{year}{2025}) \bibinfo{pages}{ep574}. \DOIprefix\doi{10.30935/cedtech/16108}.
\bibitem[{Bank(2024)}]{worldbank2024ai}
\bibinfo{author}{W.~Bank}, \bibinfo{title}{Ai: The software can instantly grade}, \bibinfo{year}{2024}. \URLprefix \url{https://documents1.worldbank.org/curated/en/099734306182493324/pdf/IDU152823b13109c514ebd19c241a289470b6902.pdf}, \bibinfo{note}{accessed: 2025-05-25}.
\bibitem[{Ravšelj(2025)}]{ravseljHigherEducationStudents2025}
\bibinfo{author}{D.~e.~a. Ravšelj},
\newblock \bibinfo{title}{Higher education students’ perceptions of {{ChatGPT}}: {{A}} global study of early reactions},
\newblock \bibinfo{journal}{PLOS ONE} \bibinfo{volume}{20} (\bibinfo{year}{2025}) \bibinfo{pages}{e0315011}. \URLprefix \url{https://journals.plos.org/plosone/article?id=10.1371/journal.pone.0315011}. \DOIprefix\doi{10.1371/journal.pone.0315011}.
\bibitem[{Gerlich(2025)}]{gerlichAIToolsSociety2025a}
\bibinfo{author}{M.~Gerlich},
\newblock \bibinfo{title}{{{AI Tools}} in {{Society}}: {{Impacts}} on {{Cognitive Offloading}} and the {{Future}} of {{Critical Thinking}}},
\newblock \bibinfo{journal}{MDPI}  (\bibinfo{year}{2025}).
\bibitem[{Darvishi et~al.(2024)Darvishi, Khosravi, Sadiq, Ga{\v{s}}evi{\'c}, and Siemens}]{darvishi2024impact}
\bibinfo{author}{A.~Darvishi}, \bibinfo{author}{H.~Khosravi}, \bibinfo{author}{S.~Sadiq}, \bibinfo{author}{D.~Ga{\v{s}}evi{\'c}}, \bibinfo{author}{G.~Siemens},
\newblock \bibinfo{title}{Impact of ai assistance on student agency},
\newblock \bibinfo{journal}{Computers \& Education} \bibinfo{volume}{210} (\bibinfo{year}{2024}) \bibinfo{pages}{104967}.
\bibitem[{Roe and Perkins(2024)}]{roe2024generative}
\bibinfo{author}{J.~Roe}, \bibinfo{author}{M.~Perkins},
\newblock \bibinfo{title}{Generative ai and agency in education: A critical scoping review and thematic analysis},
\newblock \bibinfo{journal}{arXiv preprint arXiv:2411.00631}  (\bibinfo{year}{2024}).
\bibitem[{Rodríguez-Ruiz et~al.(2025)Rodríguez-Ruiz, Marín-López, and Espejo-Siles}]{rodriguez-ruizArtificialIntelligenceUse2025}
\bibinfo{author}{J.~Rodríguez-Ruiz}, \bibinfo{author}{I.~Marín-López}, \bibinfo{author}{R.~Espejo-Siles},
\newblock \bibinfo{title}{Is artificial intelligence use related to self-control, self-esteem and self-efficacy among university students?},
\newblock \bibinfo{journal}{Education and Information Technologies} \bibinfo{volume}{30} (\bibinfo{year}{2025}) \bibinfo{pages}{2507--2524}. \URLprefix \url{https://doi.org/10.1007/s10639-024-12906-6}. \DOIprefix\doi{10.1007/s10639-024-12906-6}.
\bibitem[{Kwapisz et~al.(2024)Kwapisz, Kohli, and Rajivan}]{kwapiszPrivacyConcernsStudent2024}
\bibinfo{author}{M.~B. Kwapisz}, \bibinfo{author}{A.~Kohli}, \bibinfo{author}{P.~Rajivan},
\newblock \bibinfo{title}{Privacy {{Concerns}} of {{Student Data Shared}} with {{Instructors}} in an {{Online Learning Management System}}},
\newblock in: \bibinfo{booktitle}{Proceedings of the 2024 {{CHI Conference}} on {{Human Factors}} in {{Computing Systems}}}, {{CHI}} '24, \bibinfo{publisher}{Association for Computing Machinery}, \bibinfo{year}{2024}, pp. \bibinfo{pages}{1--16}. \URLprefix \url{https://dl.acm.org/doi/10.1145/3613904.3642914}. \DOIprefix\doi{10.1145/3613904.3642914}.
\bibitem[{Zhai et~al.(2024)Zhai, Wibowo, and Li}]{zhaiEffectsOverrelianceAI2024}
\bibinfo{author}{C.~Zhai}, \bibinfo{author}{S.~Wibowo}, \bibinfo{author}{L.~D. Li},
\newblock \bibinfo{title}{The effects of over-reliance on ai dialogue systems on students' cognitive abilities: a systematic review},
\newblock \bibinfo{journal}{Smart Learning Environments} \bibinfo{volume}{11} (\bibinfo{year}{2024}) \bibinfo{pages}{28}.
\bibitem[{Williams(2024)}]{williams2024ethical}
\bibinfo{author}{R.~T. Williams},
\newblock \bibinfo{title}{The ethical implications of using generative chatbots in higher education},
\newblock in: \bibinfo{booktitle}{Frontiers in Education}, volume~\bibinfo{volume}{8}, \bibinfo{organization}{Frontiers Media SA}, \bibinfo{year}{2024}, p. \bibinfo{pages}{1331607}.
\bibitem[{Zhang et~al.(2024)Zhang, Zhao, Zhou, and Kim}]{zhangYouHaveAI2024}
\bibinfo{author}{S.~Zhang}, \bibinfo{author}{X.~Zhao}, \bibinfo{author}{T.~Zhou}, \bibinfo{author}{J.~H. Kim},
\newblock \bibinfo{title}{Do you have {{AI}} dependency? {{The}} roles of academic self-efficacy, academic stress, and performance expectations on problematic {{AI}} usage behavior},
\newblock \bibinfo{journal}{International Journal of Educational Technology in Higher Education} \bibinfo{volume}{21} (\bibinfo{year}{2024}) \bibinfo{pages}{34}. \URLprefix \url{https://doi.org/10.1186/s41239-024-00467-0}. \DOIprefix\doi{10.1186/s41239-024-00467-0}.
\bibitem[{Risko and Gilbert(2016)}]{riskoCognitiveOffloading2016}
\bibinfo{author}{E.~F. Risko}, \bibinfo{author}{S.~J. Gilbert},
\newblock \bibinfo{title}{Cognitive {{Offloading}}},
\newblock \bibinfo{journal}{Trends in Cognitive Sciences} \bibinfo{volume}{20} (\bibinfo{year}{2016}) \bibinfo{pages}{676--688}. \URLprefix \url{https://www.cell.com/trends/cognitive-sciences/abstract/S1364-6613(16)30098-5}. \DOIprefix\doi{10.1016/j.tics.2016.07.002}. \href{http://arxiv.org/abs/27542527}{{\tt arXiv:27542527}}.
\bibitem[{Halpern(2010)}]{halpern2010halpern}
\bibinfo{author}{D.~F. Halpern},
\newblock \bibinfo{title}{The halpern critical thinking assessment: Manual},
\newblock \bibinfo{journal}{Modling, Austria: Schuhfried GmbH}  (\bibinfo{year}{2010}).
\bibitem[{Mezirow(1997)}]{mezirowTransformativeLearningTheory1997}
\bibinfo{author}{J.~Mezirow},
\newblock \bibinfo{title}{Transformative {{Learning}}: {{Theory}} to {{Practice}}},
\newblock \bibinfo{journal}{New Directions for Adult and Continuing Education} \bibinfo{volume}{1997} (\bibinfo{year}{1997}) \bibinfo{pages}{5--12}. \URLprefix \url{https://onlinelibrary.wiley.com/doi/10.1002/ace.7401}. \DOIprefix\doi{10.1002/ace.7401}.
\bibitem[{Freeman et~al.(2014)Freeman, Eddy, McDonough, Smith, Okoroafor, Jordt, and Wenderoth}]{freeman2014active}
\bibinfo{author}{S.~Freeman}, \bibinfo{author}{S.~L. Eddy}, \bibinfo{author}{M.~McDonough}, \bibinfo{author}{M.~K. Smith}, \bibinfo{author}{N.~Okoroafor}, \bibinfo{author}{H.~Jordt}, \bibinfo{author}{M.~P. Wenderoth},
\newblock \bibinfo{title}{Active learning increases student performance in science, engineering, and mathematics},
\newblock \bibinfo{journal}{Proceedings of the national academy of sciences} \bibinfo{volume}{111} (\bibinfo{year}{2014}) \bibinfo{pages}{8410--8415}.
\bibitem[{Fosnot(2013)}]{fosnot2013constructivism}
\bibinfo{author}{C.~T. Fosnot}, \bibinfo{title}{Constructivism: Theory, perspectives, and practice}, \bibinfo{publisher}{Teachers College Press}, \bibinfo{year}{2013}.
\bibitem[{Van Der~Stuyf(2002)}]{van2002scaffolding}
\bibinfo{author}{R.~R. Van Der~Stuyf},
\newblock \bibinfo{title}{Scaffolding as a teaching strategy},
\newblock \bibinfo{journal}{Adolescent learning and development} \bibinfo{volume}{52} (\bibinfo{year}{2002}) \bibinfo{pages}{5--18}.
\bibitem[{{OpenAI}(2025)}]{openai2025aiready}
\bibinfo{author}{{OpenAI}}, \bibinfo{title}{Building an {AI}{-}Ready Workforce: A Look at College Student ChatGPT Adoption in the US}, \bibinfo{type}{Technical Report}, OpenAI, \bibinfo{year}{2025}. \URLprefix \url{https://cdn.openai.com/global-affairs/openai-edu-ai-ready-workforce.pdf}, \bibinfo{note}{online; January 2025}.
\bibitem[{Kestin* et~al.(2024)Kestin*, Miller*, Klales, Milbourne, and Ponti}]{kestin*AITutoringOutperforms2024}
\bibinfo{author}{G.~Kestin*}, \bibinfo{author}{K.~Miller*}, \bibinfo{author}{A.~Klales}, \bibinfo{author}{T.~Milbourne}, \bibinfo{author}{G.~Ponti}, \bibinfo{title}{{{AI Tutoring Outperforms Active Learning}}}, \bibinfo{year}{2024}. \URLprefix \url{https://www.researchsquare.com/article/rs-4243877/v1}. \DOIprefix\doi{10.21203/rs.3.rs-4243877/v1}.
\bibitem[{Al-Zahrani(2024)}]{al-zahraniUnveilingShadowsHype2024}
\bibinfo{author}{A.~M. Al-Zahrani},
\newblock \bibinfo{title}{Unveiling the shadows: {{Beyond}} the hype of {{AI}} in education},
\newblock \bibinfo{journal}{Heliyon} \bibinfo{volume}{10} (\bibinfo{year}{2024}). \URLprefix \url{https://www.cell.com/heliyon/abstract/S2405-8440(24)06727-6}. \DOIprefix\doi{10.1016/j.heliyon.2024.e30696}.
\bibitem[{Bjork et~al.(2011)Bjork, Bjork et~al.}]{bjork2011making}
\bibinfo{author}{E.~L. Bjork}, \bibinfo{author}{R.~A. Bjork}, et~al.,
\newblock \bibinfo{title}{Making things hard on yourself, but in a good way: Creating desirable difficulties to enhance learning},
\newblock \bibinfo{journal}{Psychology and the real world: Essays illustrating fundamental contributions to society} \bibinfo{volume}{2} (\bibinfo{year}{2011}).
\bibitem[{de~Bruin et~al.(2023)de~Bruin, Biwer, Hui, Onan, David, and Wiradhany}]{de2023worth}
\bibinfo{author}{A.~B. de~Bruin}, \bibinfo{author}{F.~Biwer}, \bibinfo{author}{L.~Hui}, \bibinfo{author}{E.~Onan}, \bibinfo{author}{L.~David}, \bibinfo{author}{W.~Wiradhany},
\newblock \bibinfo{title}{Worth the effort: The start and stick to desirable difficulties (s2d2) framework},
\newblock \bibinfo{journal}{Educational Psychology Review} \bibinfo{volume}{35} (\bibinfo{year}{2023}) \bibinfo{pages}{41}.
\bibitem[{Mera et~al.(2022)Mera, Rodríguez, and Marin-Garcia}]{meraUnravelingBenefitsExperiencing2022}
\bibinfo{author}{Y.~Mera}, \bibinfo{author}{G.~Rodríguez}, \bibinfo{author}{E.~Marin-Garcia},
\newblock \bibinfo{title}{Unraveling the benefits of experiencing errors during learning: {{Definition}}, modulating factors, and explanatory theories},
\newblock \bibinfo{journal}{Psychonomic Bulletin \& Review} \bibinfo{volume}{29} (\bibinfo{year}{2022}) \bibinfo{pages}{753--765}. \URLprefix \url{https://doi.org/10.3758/s13423-021-02022-8}. \DOIprefix\doi{10.3758/s13423-021-02022-8}.
\bibitem[{Deslauriers et~al.(2019)Deslauriers, McCarty, Miller, Callaghan, and Kestin}]{deslauriersMeasuringActualLearning2019}
\bibinfo{author}{L.~Deslauriers}, \bibinfo{author}{L.~S. McCarty}, \bibinfo{author}{K.~Miller}, \bibinfo{author}{K.~Callaghan}, \bibinfo{author}{G.~Kestin},
\newblock \bibinfo{title}{Measuring actual learning versus feeling of learning in response to being actively engaged in the classroom},
\newblock \bibinfo{journal}{Proceedings of the National Academy of Sciences} \bibinfo{volume}{116} (\bibinfo{year}{2019}) \bibinfo{pages}{19251--19257}. \URLprefix \url{https://www.pnas.org/doi/abs/10.1073/pnas.1821936116}. \DOIprefix\doi{10.1073/pnas.1821936116}.
\bibitem[{Biwer et~al.(2020)Biwer, de~Bruin, Schreurs, and oude Egbrink}]{biwer2020future}
\bibinfo{author}{F.~Biwer}, \bibinfo{author}{A.~B. de~Bruin}, \bibinfo{author}{S.~Schreurs}, \bibinfo{author}{M.~G. oude Egbrink},
\newblock \bibinfo{title}{Future steps in teaching desirably difficult learning strategies: Reflections from the study smart program},
\newblock \bibinfo{journal}{Journal of Applied Research in Memory and Cognition} \bibinfo{volume}{9} (\bibinfo{year}{2020}) \bibinfo{pages}{439--446}.
\bibitem[{Howlader et~al.(2025)Howlader, Mortuja~Mahamud, Sayeeda, Sanjoy~Kumar, and Guo}]{howladerFactorsInfluencingAcceptance2025}
\bibinfo{author}{M.~H. Howlader}, \bibinfo{author}{T.~Mortuja~Mahamud}, \bibinfo{author}{Z.~Sayeeda}, \bibinfo{author}{C.~Sanjoy~Kumar}, \bibinfo{author}{J.~Guo},
\newblock \bibinfo{title}{Factors influencing the acceptance and usage of {{ChatGPT}} as an emerging learning tool among higher education students in {{Bangladesh}}: A structural equation modeling},
\newblock \bibinfo{journal}{Cogent Education} \bibinfo{volume}{12} (\bibinfo{year}{2025}) \bibinfo{pages}{2504224}. \URLprefix \url{https://doi.org/10.1080/2331186X.2025.2504224}. \DOIprefix\doi{10.1080/2331186X.2025.2504224}.
\bibitem[{Klimova and Pikhart(2025)}]{klimova2025exploring}
\bibinfo{author}{B.~Klimova}, \bibinfo{author}{M.~Pikhart},
\newblock \bibinfo{title}{Exploring the effects of artificial intelligence on student and academic well-being in higher education: a mini-review},
\newblock \bibinfo{journal}{Frontiers in Psychology} \bibinfo{volume}{16} (\bibinfo{year}{2025}) \bibinfo{pages}{1498132}.
\bibitem[{Be{\'c}irovi{\'c} et~al.(2025)Be{\'c}irovi{\'c}, Polz, and Tinkel}]{becirovic2025exploring}
\bibinfo{author}{S.~Be{\'c}irovi{\'c}}, \bibinfo{author}{E.~Polz}, \bibinfo{author}{I.~Tinkel},
\newblock \bibinfo{title}{Exploring students’ ai literacy and its effects on their ai output quality, self-efficacy, and academic performance},
\newblock \bibinfo{journal}{Smart Learning Environments} \bibinfo{volume}{12} (\bibinfo{year}{2025}) \bibinfo{pages}{29}.
\bibitem[{Gerlich(2023)}]{gerlichPowerVirtualInfluencers2023}
\bibinfo{author}{M.~Gerlich},
\newblock \bibinfo{title}{The {{Power}} of {{Virtual Influencers}}: {{Impact}} on {{Consumer Behaviour}} and {{Attitudes}} in the {{Age}} of {{AI}}},
\newblock \bibinfo{journal}{Administrative Sciences} \bibinfo{volume}{13} (\bibinfo{year}{2023}) \bibinfo{pages}{178}. \URLprefix \url{https://www.mdpi.com/2076-3387/13/8/178}. \DOIprefix\doi{10.3390/admsci13080178}.
\bibitem[{Schaaff and Heidelmann(2024)}]{schaaff2024impacts}
\bibinfo{author}{K.~Schaaff}, \bibinfo{author}{M.-A. Heidelmann},
\newblock \bibinfo{title}{Impacts of anthropomorphizing large language models in learning environments},
\newblock \bibinfo{journal}{arXiv preprint arXiv:2408.03945}  (\bibinfo{year}{2024}).
\bibitem[{Bastani et~al.(2024)Bastani, Bastani, Sungu, Ge, Kabakci, and Mariman}]{bastaniGenerativeAICan2024}
\bibinfo{author}{H.~Bastani}, \bibinfo{author}{O.~Bastani}, \bibinfo{author}{A.~Sungu}, \bibinfo{author}{H.~Ge}, \bibinfo{author}{O.~Kabakci}, \bibinfo{author}{R.~Mariman}, \bibinfo{title}{Generative {{AI Can Harm Learning}}}, \bibinfo{year}{2024}. \URLprefix \url{https://www.ssrn.com/abstract=4895486}. \DOIprefix\doi{10.2139/ssrn.4895486}.
\bibitem[{Chan(2024)}]{chanExploringFactorsAI2024}
\bibinfo{author}{C.~K.~Y. Chan}, \bibinfo{title}{Exploring the {{Factors}} of "{{AI Guilt}}" {{Among Students}} -- {{Are You Guilty}} of {{Using AI}} in {{Your Homework}}?}, \bibinfo{year}{2024}. \URLprefix \url{http://arxiv.org/abs/2407.10777}. \DOIprefix\doi{10.48550/arXiv.2407.10777}. \href{http://arxiv.org/abs/2407.10777}{{\tt arXiv:2407.10777}}.
\bibitem[{Elliot and Devine(1994)}]{elliot1994motivational}
\bibinfo{author}{A.~J. Elliot}, \bibinfo{author}{P.~G. Devine},
\newblock \bibinfo{title}{On the motivational nature of cognitive dissonance: Dissonance as psychological discomfort.},
\newblock \bibinfo{journal}{Journal of personality and social psychology} \bibinfo{volume}{67} (\bibinfo{year}{1994}) \bibinfo{pages}{382}.
\bibitem[{Shores(2024)}]{shores2024surveillance}
\bibinfo{author}{T.~Shores}, \bibinfo{title}{Surveillance Education: Tracks the Rise of Spying Technology in Schools}, \bibinfo{publisher}{MIT Press}, \bibinfo{year}{2024}.
\bibitem[{Mason et~al.(2016)Mason, Yerushalmi, Cohen, and Singh}]{mason2016learning}
\bibinfo{author}{A.~Mason}, \bibinfo{author}{E.~Yerushalmi}, \bibinfo{author}{E.~Cohen}, \bibinfo{author}{C.~Singh},
\newblock \bibinfo{title}{Learning from mistakes: The effect of students' written self-diagnoses on subsequent problem solving},
\newblock \bibinfo{journal}{The Physics Teacher} \bibinfo{volume}{54} (\bibinfo{year}{2016}) \bibinfo{pages}{87--90}.
\bibitem[{Niloy et~al.(2024)Niloy, Hafiz, Hossain, Gulmeher, Sultana, Islam, Bushra, Islam, Hoque, Rahman, and Kabir}]{niloyAIChatbotsDisguised2024}
\bibinfo{author}{A.~C. Niloy}, \bibinfo{author}{R.~Hafiz}, \bibinfo{author}{B.~M. Hossain}, \bibinfo{author}{F.~Gulmeher}, \bibinfo{author}{N.~Sultana}, \bibinfo{author}{K.~F. Islam}, \bibinfo{author}{F.~Bushra}, \bibinfo{author}{S.~Islam}, \bibinfo{author}{S.~I. Hoque}, \bibinfo{author}{M.~Rahman}, \bibinfo{author}{S.~Kabir},
\newblock \bibinfo{title}{{{AI}} chatbots: {{A}} disguised enemy for academic integrity?},
\newblock \bibinfo{journal}{International Journal of Educational Research Open} \bibinfo{volume}{7} (\bibinfo{year}{2024}) \bibinfo{pages}{100396}. \URLprefix \url{http://www.sciencedirect.com/science/article/pii/S2666374024000785}.
\bibitem[{{Digital Education Council}(2025)}]{digitaleducationcouncil2025}
\bibinfo{author}{{Digital Education Council}}, \bibinfo{title}{Digital education council}, \bibinfo{howpublished}{\url{https://www.digitaleducationcouncil.com}}, \bibinfo{year}{2025}. \bibinfo{note}{Accessed: 2025-05-27}.
\bibitem[{Favero et~al.(2024)Favero, P{\'e}rez-Ortiz, K{\"a}ser, and Oliver}]{favero2024enhancing}
\bibinfo{author}{L.~Favero}, \bibinfo{author}{J.~A. P{\'e}rez-Ortiz}, \bibinfo{author}{T.~K{\"a}ser}, \bibinfo{author}{N.~Oliver},
\newblock \bibinfo{title}{Enhancing critical thinking in education by means of a socratic chatbot},
\newblock \bibinfo{journal}{arXiv preprint arXiv:2409.05511}  (\bibinfo{year}{2024}).

\end{thebibliography}

\end{document}